\documentstyle[11pt,aaspp4]{article}

\def\Gyr{\,{\rm Gyr}}

\def\pc{\,{\rm pc}}

\def\mpc{\,{\rm Mpc}}

\def\kms{\,{\rm km\,s^{-1}}}

\def\beq{\begin{equation}}
\def\eeq{\end{equation}}
\def\beqar{\begin{eqnarray}}
\def\eeqar{\end{eqnarray}}

\def\msol{\hbox{$M_{\odot}$}}
\def\csfr{\hbox{$\dot{\rho}_\star$}}

\def\lsfru{\hbox{$M_\odot \, {\rm pc}^{-2} \, {\rm Gyr}^{-1}$}}

\def\mto{\hbox{$m_{0}$}}
\def\pdmf{\hbox{$\Phi_{\rm MS}$}}
\def\cden{\hbox{$\Sigma_{\star}$}}
\def\lum#1{{\cal L}_{#1}}

\def\pcite#1{(\cite{#1})}
\def\pref#1{(\ref{#1})}

\def\bline#1{\newlength{\citewidth}\settowidth{\citewidth}{#1}
  \rule[1.2mm]{\citewidth}{0.1mm}}

\def\la{\mathrel{\mathpalette\fun <}}
\def\ga{\mathrel{\mathpalette\fun >}}
\def\fun#1#2{\lower3.6pt\vbox{\baselineskip0pt\lineskip.9pt
  \ialign{$\mathsurround=0pt#1\hfil##\hfil$\crcr#2\crcr\sim\crcr}}}

\begin{document}

%\slugcomment{UMN-TH-1621/98}
%\slugcomment{in press, {\it The Astrophysical Journal}}
\rightline{UMN-TH-1621/98}
\rightline{in press, {\it The Astrophysical Journal}}

\pagestyle{myheadings}

\title{TESTING THE RELATION BETWEEN \\
THE LOCAL AND COSMIC STAR FORMATION HISTORIES}

\author{Brian D. Fields}
\affil{School of Physics and Astronomy, University of Minnesota  \\ 
Minneapolis, MN 55455, USA \\
{\tt fields@mnhepw.hep.umn.edu} }
\authoremail{fields@mnhepw.hep.umn.edu}

\begin{abstract} 
Recently, there has been great progress toward observationally
determining the mean star formation history of the universe.  
When accurately known, the cosmic star formation rate could provide much
information about Galactic evolution, if the 
Milky Way's star
formation rate is representative of the average cosmic star formation
history.  A simple hypothesis is that our local star formation rate is
proportional to the cosmic mean.  In addition, to specify a star formation
history, one must also adopt an initial mass function (IMF); typically it
is assumed that the IMF is a smooth function which is constant
in time.  We show how to test directly the compatibility of all these
assumptions, by making use of the local (solar neighborhood)
star formation record encoded
in the present-day stellar mass function.  
Present data suggests that at least one
of the following is false: (1) the local IMF is constant in time; (2)
the local IMF is a smooth (unimodal) function; and/or (3) star
formation in the Galactic disk was representative of the
cosmic mean.  We briefly discuss how to determine which of these
assumptions fail, and improvements in observations which will
sharpen this test.

\end{abstract}


\section{Introduction}
\label{sec:intro}

The cosmic star formation history has recently begun to be unveiled
and pushed to increasingly high redshift 
(Gallego et al.\ \cite{gall};
Lilly et al.\ \cite{lilly};
Madau et al.\ \cite{mad96};
Madau \cite{mad97a,mad97b};
Connolly et al.\ \cite{conn};
Pettini et al.\ \cite{pett};
Tresse \& Maddox \cite{tm}).
Current results suggest a sharp rise to a peak at
reshifts $z \sim 1-3$,
though the behavior at $z \ga 1$
depends strongly on models of
absorption at these epochs
(see \S \ref{sec:csfr}).
In addition to their cosmological importance,
these data are potentially useful for understanding the
evolution of our own Galaxy, if the Milky Way's 
star formation history has been typical.  Knowledge
of the past Galactic star formation rate would
significantly reduce uncertainties and {\it ad hoc} assumptions
in chemical evolution 
calculations.

Thus, we wish to consider
the connection between the cosmic star formation rate
(CSFR) and the 
local (i.e., solar neighborhood)
star formation rate (LSFR). 
In particular, we will test the hypothesis that 
the cosmic and local star formation rates are
simply related.  The simplest ansatz
is that the two rates are proportional to one another;
we will  refer to this
as the hypothesis of ``star formation universality.''
Whether the Galaxy's evolution was representative---i.e.,
followed the cosmic mean---is not certain.
On the one hand, the Milky Way appears to be 
typical in many respects, with a common 
morphology and a luminosity $\sim L_*$ 
near the local, low-$z$ average.
On the other hand, it is 
certain that in some systems, star formation 
can and does
proceed differently from the cosmic 
mean.  For example, starburst galaxies and extragalactic HII regions
show highly elevated
star formation rates that cannot be maintained
smoothly over cosmic timescales.
Also, a hierarchical clustering scheme
of structure formation predicts that 
smaller objects form first and later merge to make
large galaxies such as our own, with star formation
commencing in protogalaxies at redshifts
$\simeq 3.5$ (Baugh, Cole, Frenk, \& Lacey \cite{bcfl}).

In testing for star formation universality, 
an additional complication comes into play:
to fully specify a star formation history 
requires not only a star formation rate, but
also an initial mass function (IMF).
Indeed, the IMF and the LSFR are closely entangled---it is
difficult to determine the IMF over
all of the observed stellar mass range
without some knowledge the star formation history.
Nevertheless, it is typically assumed
(with support from observations: Wyse \cite{rosie};
Hillenbrand \cite{hill})
that the IMF is constant in time, and 
``smoothly varying''---i.e., unimodal in
shape; this form has some support from
theory (e.g., Adams \& Fatuzzo \cite{af}; Silk \cite{silk};
Ferrini, Palla, \& Penco \cite{fpp}).
On the other hand, some models have adopted bimodal star formation
(e.g., Larson \cite{larson}; Gusten \& Mezger \cite{gm}; 
Wyse \& Silk \cite{ws}),
motivated in part by evidence that high- and low-mass star formation in
molecular clouds does occur in spatially separated regions
(Herbig \cite{herb}).
It is not clear, however, whether this local spatial variation leads to
bimodality in the global, ensemble averaged IMF.
For example, Ferrini, Palla, \& Penco \pcite{fpp} present a
fragmentation theory for star formation in which the mass distribution
varies according to the local properties of
the molecular cloud, but the global superposition of these
distributions leads to a fairly smooth and unimodal IMF.
Given this theoretical uncertainty,
model-builders have for the most part adopted an IMF which is a single
power law, or broken power law sometimes
approximated as a lognormal.

In this paper, we will examine the implications
of star formation universality by combining the most common
assumptions about the Galactic IMF.
The compatibility of these assumptions is
testable via the present day mass function (PDMF),
i.e., the local (solar neighborhood) stellar luminosity
function, converted to a distribution in mass.
The PDMF which contains information about the IMF
as well as the LSFR over a wide range of epochs.
It is important thus to note that by using the PDMF for
our comparison of local versus cosmic star formation,
we take ``local'' star formation to mean
that in the {\em solar neighborhood}, where the PDMF
is observed.  We thus can at best hope that the
PDMF samples the disk
star forming history, and we must remember
that we have not included the history of the stellar halo.

In our test of star formation universality,
we use the observed 
PDMF to infer the needed IMF, whose
smoothness (unimodality) we examine.
We also
go the other direction, 
computing the expected PDMF for typical IMF choices,
and comparing this with the data.  In both cases,
we find that the data suggest 
that our Galaxy cannot have both 
typical evolution and a time-independent, smooth IMF.

We present the needed formalism in \S \ref{sec:form},
and summarize the 
relevant data in \S \ref{sec:csfr}.
We calculate the PDMF--IMF transformations 
in \S \ref{sec:test}, and in \S \ref{sec:local} discuss the possible 
differences between
local, Galactic, and cosmic star formation histories.
The implications of
our results are explored in \S \ref{sec:fin}.

\section{Formalism}
\label{sec:form}

The fundamental object quantifying star formation
history (Salpeter \cite{salp}; Tinsley \cite{tins})
is the stellar creation function
$C(m,t)$.  This measures the
number $N_\star$ of stars born
in mass range $(m,m+dm)$ and time interval $(t,t+dt)$,
and is defined by
\beqar
\label{eq:Cdef}
d N_\star & = & C(m,t) \ dm \, dt \\
\label{eq:Csep}
  & \stackrel{\mathrm sep}{=} & \phi(m) \ \psi(t) \ dm \, dt 
\eeqar
For local (i.e., solar neighborhood) stars, 
$N_\star$ is usually expressed a column density,
averaged over disk scale height, 
of newborn stars.  
The creation function is often 
assumed to be {\em separable}, i.e., in the form of
eq.\ \pref{eq:Csep},
where the IMF is $\phi$, and the LSFR is $\psi$.
The units for the IMF and LSFR depend on the
normalization of $\phi$; for the
customary choice of $\int dm \, m\, \phi(m) = 1$ (which we will adopt), 
$[\psi] = \lsfru$.
For a thorough review of the creation function and the IMF,
see Scalo \pcite{scalo}.

Adopting a separable creation function is
equivalent to assuming the IMF is constant in time.
Clearly, this is a very strong assumption,
one that greatly reduces the large freedom available
in a general creation function.
Nevertheless, model-builders have, whenever possible,
made this
ansatz on the basis of simplicity.
In this paper, we will adopt separability as well,
including it among the basic assumptions whose 
compatibility we wish to test.

The data we will use to test star formation histories
is the disk present day mass function, which quantifies
the mass distribution of 
main-sequence stars burning in the disk today.
Observationally, the PDMF is derived from the
luminosity function of solar neighborhood,
main-sequence stars, via a translation to
a mass function via a mass--luminosity relation.
Theoretically, the PDMF can be derived in full
generality from eq.\ \pref{eq:Cdef}.
One of two expressions is appropriate, depending
on the main sequence mass $m$ and its associated
lifetime $\tau(m)$.  
The two cases are divided
at the present-day main sequence turnoff mass $m_0$,
defined by $\tau(m_0) \equiv t_0$, where
$t_0$ is the present age of the universe.\footnote{Strictly speaking,
one should use $\tau(\mto) = t_{\rm disk}$, the age of the disk,
but in practice the difference is relevant for a very
narrow range of masses only; see the discussion in \S \ref{sec:local}.}
Specifically, we have
\beq
\label{eq:pdmf_general}
\pdmf(m) = \left\{
  \begin{array}{ll}
      \int_{0}^{t_0} \ dt^\prime \ C(m,t)
        & m < \mto \\
      \int_{t_0-\tau(m)}^{t_0} \ dt^\prime \ C(m,t) 
        & m \ge \mto \\
  \end{array}
  \right. 
\eeq
Note that, for a known creation function $C$, the PDMF
is completely determined.  Equation (\ref{eq:pdmf_general})
is thus fundamentally simpler than, e.g., the usual 
integro-differential expressions for
mass or abundance consumption in chemical evolution.
The simplicity of the connection between the star formation 
history $C$ and the (observable) PDMF is one of the strengths
of the present analysis.

In our case of a separable creation function, 
eq.\ \pref{eq:pdmf_general} specializes to 
\beq
\label{eq:pdmf}
\pdmf(m) = \left\{
  \begin{array}{ll}
   \cden(t_0) \ \phi(m) \ 
        & m < \mto \\
   \left( \cden(t_0) - \cden\left[t_0-\tau(m)\right] \right) \ \phi(m) 
        & m \ge \mto \\
  \end{array}
  \right.
\eeq
The quantity
\beq
\cden(t) \equiv \int_{0}^{t} \ dt^\prime \ \psi(t^\prime) 
\eeq
is the aggregate mass (per square parsec) ever going into stars by time $t$,
including material that went into now-dead stars.
In physical terms, eq.\ \pref{eq:pdmf} states that
at a given mass, the PDMF is given by the product of 
the IMF at that mass, times the integrated star formation over
the stellar lifetime at that mass, or the age of the Galaxy,
whichever is smaller.
Note that the PDMF has units 
$[\pdmf] = {\rm stars} \, \pc^{-2} \, \msol^{-1}$.
The observed PDMF of Scalo (\cite{scalo}) appears
as the filled points in Figures \ref{fig:pdmf_Salp} and \ref{fig:pdmf_MS}.

As is well-known, \pdmf\ simplifies in 
two limits.  For high masses ($m \ga 4-5 \msol$), 
the lifetimes are short enough ($\tau(m) \ll t_0$), 
that one essentially samples only the present
LSFR:
\beq
\label{eq:himass}
\pdmf(m) \approx \phi(m) \, \tau(m) \, \psi(t_0)
\eeq
Since $\tau(m)$ is known theoretically to fairly good
accuracy, the shape of $\phi(m)$ is easily inferred
from $\pdmf$ in this mass range.
The PDMF also simplifies at low masses. 
For a present age $t_0 = 13$ Gyr (15 Gyr),
the turnoff is at
$\mto = 0.91 \msol$ ($0.87 \msol$).
At low masses  $m < \mto$, we have
$\pdmf(m,t_0) = \cden(t_0) \, \phi(m)$.
Again, in this range, the PDMF gives the shape of the IMF,
though here the scaling is different than for high masses,
now depending on the aggregate star formation 
(or, equivalently, the time averaged star formation).
Indeed, the usual procedure for deriving the IMF
from the PDMF is to smoothly match the IMF derived
from high and low mass limits (for an assumed Galactic
star formation rate).  The matching gives the ratio 
$\tau_\star = \cden/\psi$, a measure of the
timescale for star formation.

Clearly, the PDMF directly expresses information about star formation
history as encoded in both the IMF and LSFR.  Unfortunately,
the relation is not a uniquely invertible one:  the PDMF
can be consistent with many putative star formation histories,
even in the case of a separable creation function.  However,
the PDMF is certainly {\it not} consistent will {\it all}
possible histories.  As first noted by Salpeter \pcite{salp}, 
and emphasized by Scalo \pcite{scalo},
one must show that any choice of IMF and LSFR are
{\it compatible}; indeed tests of IMF--PDMF compatibility have become
commonplace features of chemical
evolution studies (as reviewed in e.g., Tinsley \cite{tins};
Shore \& Ferrini \cite{sf}; or Pagel \cite{pagel}). 
One approach to testing the PDMF---star formation history compatibility 
is to adopt both a LSFR and an IMF, and to those that these 
are able to reproduce the PDMF
via eq.\ \pref{eq:pdmf}.  Alternatively, one can choose
to adopt either an IMF or a LSFR, and via eq.\ \pref{eq:pdmf} use the
PDMF to deduce the other function; this has been done with particular
care in 
Mathews, Bazan, \& Cowan \pcite{mbc}.  
In this paper we will use both procedures to address the
compatibility of typical assumptions about the IMF
with the emerging data on the cosmic star formation
rate, as described in the following section.

\section{The Cosmic Star Formation Rate}
\label{sec:csfr}

The CSFR, which we will denote $\csfr$,
is the average cosmic rate at which mass goes into stars 
per unit comoving volume.
It is obtained via analysis of the
galactic luminosity function at various
epochs and over a range of (comoving) wavelengths.  Of particular interest
is the redshift history of 
the integrated galactic 
emission from hot, short-lived stars,
either from the H$\alpha$ line, or from the UV continuum.
Because essentially all H$\alpha$ or UV comes from
these massive stars, it follows that
the luminosity density at these wavelengths 
(when corrected for absorption)
is proportional to the instantaneous CSFR
(e.g., Kennicutt \cite{ken}; Madau et al. \cite{mad96}).
Strictly speaking, the UV or H$\alpha$ luminosity 
traces only the {\em massive} star formation, but
if the IMF is time-independent as we assume, then
the massive star formation rate differs from the total
star formation rate by just a constant factor.

Furthermore, the {\it shape} of the CSFR inferred from
massive star luminosity
is independent of
the cosmic IMF;
this follows because there is, for any reasonable IMF,
only a very small contribution to these wavelengths
from lower mass, longer lived stars.
However, the {\it normalization} of the CSFR
does depend on the IMF, which quantifies the
fraction of all stars which are massive.
Fortunately, we will only need the shape of the
CSFR, as we only wish to test the ansatz that
$\psi \propto \csfr$.

For $z \la 1$, Lilly et al.\ (\cite{lilly})
used the Canada-France-Hawaii survey data
to compute the evolution 
of the comoving UV luminosity density at $\lambda = 2800$ \AA,
$\lum{2800}$, which implies
\beq
\csfr(z) \propto \lum{2800}(z) \propto (1+z)^{3.9\pm0.75} 
\, , \ \ {\rm for} \ z \la 1 \ 
\eeq
(throughout, densities are expressed in comoving units).
For comparison, 
Gallego et al.\ \pcite{gall}
and Tresse \& Maddox \pcite{tm} use 
H$\alpha$ luminosity as a massive star formation
indicator.  Tresse \& Maddox thereby find a somewhat higher increase
with redshift:
$\csfr(z) \propto \lum{{\rm H}\alpha} \propto (1+z)^{4.4}$.

For redshifts $z>1$, analysis of the Hubble Deep Field 
(Madau et al.\ \cite{mad96}; Madau \cite{mad97a,mad97b},
Connolly, Szalay, Dickinson, 
SubbaRao, \& Brunner \cite{conn}) suggest
a peak in the $z=1-2$ interval, and  then
(in the case of constant extinction)
a dropoff by $z \sim 4$ to values close to those at 
$z=0$.
However, these results depend strongly on 
the absorption of the UV light as it travels to us and is
redshifted; it has been suggested that dust extinction at high
redshifts
could require upward corrections in the CSFR by factors 
of 3 (Pettini et al.\ \cite{pett}) or even more
(Meurer et al.\ \cite{meurer}).  

Madau et al.\ \pcite{mad96} were the first to use
the ``UV dropout'' analysis of the Hubble Deep Field
to construct a cosmic star formation history.
In an attempt to span the different possibilities
for cosmic absorption, 
Madau et al., and later Madau \pcite{mad97a,mad97b},
examine two models for extinction, motivated
by two galaxy formation scenarios.  
(1) Madau's fiducial model 
assumes dust extinction does not evolve with redshift, 
and takes the form of
an SMC extinction law: 
$E_{B-V}(z) = const = 0.06$ mag. 
As the extinction is not
a function of time or redshift, this model does not give
a strong correction to the rest-frame UV light to infer the
CSFR at early epochs.
(2) To try to bracket the effect of extinction on masking star formation, 
Madau also presents a model in which it is assumed
that the star formation is large at $z>1$, but 
that half of all stars born are shrouded in dust.  This model 
has extinction which increases with
redshift:  $E_{B-V}(z) = 0.0067 (1+z)^{2.2}$ mag; 
it thus leads to much larger CSFR at high redshift than does
the fiducial, constant $E_{B-V}$ model.  
The CSFRs derived from each method 
(denoted the ``constant'' and ``evolving'' $E_{B-V}$ models) appear
in the insets to the Figures.
We will use both models in our analysis.

Note the different units in the local and cosmic star formation
rates, of area and volume densities, respectively.
Physically, this comes about since
the local, disk data averages over 
scale heights, whereas the 
cosmic rate averages
over all comoving star forming material.
The different units point up the assumptions 
made in assuming that the LSFR and CSFR are proportional.
Namely, 
(1)  disk scale height has not changed over the disk lifetime,
and 
(2)  external perturbations to our own disk star formation (e.g.,
merging or infall) are typical of the average galaxy.

Note also that the CSFR results depend on adopted cosmology, 
which gives the age of universe $t_0$, and
supplies the $t-z$ relation needed to convert CSFR information
known as a function of redshift.
In this paper, we use $H_0 = 50 \kms \mpc^{-1}$,
and an $\Omega_0 = 1$ Einstein-de Sitter universe
with no cosmological constant;
this gives $t_0 = 13$ Gyr.

\section{PDMF Tests of Star Formation Universality}
\label{sec:test}

With the CSFR data, and some general assumptions
about the IMF, we may now use eq.\ \pref{eq:pdmf}.
to test star formation universality, i.e., whether 
cosmic star formation and local star formation are proportional.
Other chemical evolution
tests are possible, but these also involve IMF assumptions
and perhaps other stellar model uncertainties too,
e.g., yields.  Here the assumptions are fewer and
more explicit.   

As discussed in the previous section,
the CSFR becomes uncertain 
around $z_{\rm max} \sim 1$, 
due to poorly known extinction at high redshift;
this epoch occurs at 
$t_{\rm min} \sim 5$ Gyr for our adopted cosmology. 
The associated lookback time $t_{\rm look} = t_0 - t_{\rm max} \sim 8$ Gyr
corresponds to the lifetime of a star 
of mass $m_{\rm min} \simeq 1.1 \msol$.
Fortunately, 
there is only small gap between this 
and the turnoff mass $\mto \simeq 0.9 \msol$
Thus we can reliably reconstruct the PDMF down to $m_{\rm min}$,
but below this mass, uncertainty accumulates due to the poorly
know high-redshift CSFR.  
It is worth noting that estimates of disk age give about
$\tau_{\rm disk} \sim 10 \Gyr$, i.e., a disk birth at 
$t_0 - \tau_{\rm disk} = 3$ Gyr, which is
of order $t_{\rm min}$.
Thus, where the CSFR data becomes less certain,
we also expect some halo contribution to the CSFR, and
for the physics of disk formation to become 
a potential complication (see \S \ref{sec:local}).

As discussed in \S \ref{sec:form}, adopting a particular star formation
rate uniquely determines
the IMF (in the context of a separable creation function)
by inverting eq.\ \pref{eq:pdmf} via
\beq
\label{eq:imf}
\phi(m) \propto \left\{
  \begin{array}{ll}
     \frac{\pdmf(m)}{\cden(t_0)} 
          & m < \mto \\
     \vphantom{------} \\
     \frac{\pdmf(m)}{\cden(t_0) - \cden\left[t_0-\tau(m)\right]} 
          & m \ge \mto \\
  \end{array}
  \right.
\eeq
Figure \ref{fig:imf} plots the IMFs that result from
the Madau fiducial, ``constant $E_{B-V}$'' CSFR.
We see that the IMF is bimodal,
with distinct peaks at $m \sim 0.2 \msol$ and $m \sim 1.3 \msol$.
This appearance of
multiple peaks arises due to the strong rise of
the CSFR back towards early epochs.  That is,
since $\phi$
varys inversely with the integrated star formation
according to eq.\ \pref{eq:imf},
the long-lived, low mass end of the IMF is suppressed
relative to the high mass end.  This correction, superimposed
upon the singled peaked PDMF, leads to a double-peaked IMF.
Furthermore, note that the bimodality has a suspicious onset.
The high mass, power law trend turns over at $m \sim 1.4 \msol$,
precisely the mass where $\tau(m) \sim$ Gyr becomes comparable
to the shortest (high--$z$) input timescales.  
That is, the departure from smoothness happens abruptly, and 
at a mass scale that is neither expected to be special,
nor seen to be remarkable
in the PDMF data, and is likely to be 
an artifact of the input star formation rate.

To give a rough sense of the departure from unimodality,
we fit the IMF of Figure \ref{fig:imf}(a)
to the simplest nonlinear unimodal form,
a lognormal (i.e., quadratic in $\log \phi$ versus $\log m$).  
We find the best quadratic fit by minimization of $\chi^2$;
results appear in the upper panel of
Figure \ref{fig:quad_fit}.  For this plot,
we have computed the formal reduced $\chi^2$ for the
$\log \phi$--$\log \phi_{\rm fit}$ data versus the fit curve.  
We find $\chi^2_\nu = 0.40$, which is quite small.
Nominally, this would indicate a good fit---indeed, it suggests
that the errors are overestimated.  However, this is not
the whole story, as the $\chi^2_\nu$ value alone only quantifies
how well the ensemble of points fits the curve; it is silent
as to the distribution of points about the curve.  
In fact, it is clear from the figure that the
errors are not distributed about the best-fit curve in
a random fashion.  Instead, the data for the most part lie quite
comfortably within the curve--except around $m \sim 2 \msol$,
where the data systematically rises and then falls.
For the underlying IMF to be unimodal and lie
on the curve requires a systematic conspiracy of errors
right around $m \sim 2 \msol$ to counter the departures
seen in Figure \ref{fig:quad_fit}(b).
We therefore 
view the IMF's systematic departure from smoothness as
quite suggestive, if not yet definitive.

Thus we see that the IMF required by star formation
universality is probably not unimodal.  This is the main result
of this paper, with significant implications.  
Namely, we are led to
conclude that either: (1) our calculation
is correct, and the universal IMF has this form,
contrary to observational evidence,
theoretical prejudice, and common usage,
which favor a smoother form, either unimodal or power law;
(2) our calculation is incorrect because the IMF
is not constant in time, again contrary to common assumptions;
or (3) our calculation is incorrect because
star formation universality does not hold---the
Galactic star formation rate is not representative
of the cosmos.  
Thus, the simplest assumptions about the IMF
are inconsistent with the simplest ansatz about 
the local--cosmic star formation connection.

Our calculation does not tell which of these possibilities
is the right one; 
in the final section we discuss further
tests to address this issue.  One implication is worth
noting here:  if the IMF indeed varys with time, it
must be top-heavy---i.e., biased towards high masses---at
early times.  This follows from the mismatch between the
high-and low-mass regions of the IMF.  The low mass end
is overly suppressed because the integrated star formation
$\cden(t_0)$ is high, and $\phi \propto \cden(t_0)^{-1}$ in this
mass range.  A top-heavy early IMF reduces this suppression
by reducing the high-redshift contribution to $\cden(t_0)$
in low mass stars.  It is interesting to note that 
IMFs of this character have been discussed in the literature;
see Cass\'{e}, Olive, Vangioni-Flam, \& Audouze \pcite{cova} 
and references therein.

The uncertainties in the IMF of Figure \ref{fig:imf}
arise in part from the input PDMF, but 
more importantly from the 
(as yet) poorly known nature of cosmic star formation
(cf. \S \ref{sec:csfr}).
In Figure \ref{fig:alt_imf}, we compare the IMF of Figure \ref{fig:imf}
with one derived from the Madau ``evolving $E_{B-V}$'' scenario.
Here we see that the bimodality is even stronger, a consequence of
this scenario's higher star formation at early epochs.
The open squares of 
figure \ref{fig:alt_imf} show the IMF resulting from
using a ``truncated'' CSFR, which we take to be
the Madau CSFR for $z<1$, and zero for $z \ge 1$.
By comparing the IMF resulting from this CSFR with the IMFs from
the other two, we see the pivotal role of the evolution 
in the uncertain $z\ge1$ regime.
With the truncated CSFR, the
IMF is now completely consistent with being unimodal; 
we thus infer that the
high-redshift behavior is crucial in
forcing a bimodal IMF and the consequences thereby implied.
Improved determination of the CSFR in this redshift
range will thus sharpen our test.

For comparison, we have also considered the alternate test for
PDMF--star formation history compatibility, 
in which one
assumes a particular form of the IMF as well as the LSFR.  
If the IMF were ``known,'' it would make for a different
and more decisive  test
of star formation universality:
one derives the (fully determined) PDMF
according to eq.\ \pref{eq:pdmf}, and compares it directly to the data.
Such tests are routinely performed in chemical evolution models
(see Pagel \cite{pagel}); 
here, however, we do not require an {\it ad hoc} form
of the LSFR, but rather star formation universality.  

We will examine two commonly used IMF functional forms:
(1) a power law,  $\phi(m) \propto m^{-x}$,
with $x = 2.35$ (Salpeter \cite{salp}); and 
(2) a lognormal 
$\phi \propto m^{-1} \, \exp[-\ln^2(m/m_c)/2\sigma^2]$,
with $m_c = 0.087 \msol$, $\sigma = 1.6$ (Miller \& Scalo \cite{ms}).
These forms have
some support from work on the theory of the IMF 
(Adams \& Fatuzzo \cite{af}; Silk \cite{silk}).
However, we caution that these ``standard'' IMFs have arisen
because they provide good fits to the PDMF 
given various {\em assumed}, 
mildly varying LSFR trends.  That is, these
IMFs were not derived to allow for strongly variable star
formation, as is observed in the CSFR.
Nevertheless, we include them here for comparison.

Our results appear in Figures \ref{fig:pdmf_Salp} and \ref{fig:pdmf_MS}.  
We fix 
the derived PDMF to the data by adjusting the
(arbitrary) normalization to minimize the $\chi^2$ of the fit to the data.
We see that when one adopts either the Salpeter or the Miller-Scalo IMF,
the derived PDMF fails to agree with the data
at low ($m \la 0.2 \msol$) and intermediate ($m \sim 2 \msol$) masses.  
The discrepancy below $0.2 \msol$ arises due to the high star formation
at early times, which leads to a large integrated
star formation $\cden(t_0)$, and in turn a high
$\pdmf \propto \cden(t_0)$ at $m < \mto$.
The discrepancy around $2 \msol$ arises for similar reasons
as the IMF bimodality seen in Figures \ref{fig:imf}--\ref{fig:alt_imf},
discussed above.
Indeed, this behavior is as expected:
we have adopted a time-independent, unimodal IMF,
so we should not expect the LSFR to trace the cosmic mean.

\section{Which Local Star Formation History?}
\label{sec:local}

In considering whether the our local star formation
history is representative of the cosmic mean,
one should be clear about the possible distinctions
between the solar neighborhood, Galactic, and cosmic
star formation histories.  
As noted in \S\ref{sec:intro}, 
the PDMF encodes only star formation in the disk 
(Population I),
and omits halo (Population II) star formation.  
Furthermore, the dynamics and composition of the halo
stars strongly suggests that they were formed prior
to most disk stars, so that their contribution to the
Galactic star formation rate (globally and locally)
changes the overall {\em shape} of the time history.
Consequently, the ``local'' star formation history sampled
by the PDMF is necessarily an incomplete account of the
Galactic star formation history.
Thus, halo star formation at early epochs probably
contributes to the CSFR, but not to the PDMF,
and could thus be responsible for part of the
bimodality seen in Figures \ref{fig:imf}
and \ref{fig:alt_imf}.

We can estimate the effect of halo star formation as follows.
As we have seen, the discontinuity 
in the derived IMF
arises because the low-mass ($m \le \mto \sim 0.9 \msol$)
points suffer a large correction, because their cumulative number
includes all star formation epochs, including the earliest ones.
Specifically, eq.\ \pref{eq:imf} shows that
for low mass stars, $\phi \propto \cden(t_0)^{-1}$,
that is, the suppression is inversely proportional to the
{\em total}, integrated star formation.
In our test, we obtained $\cden$ from the CSFR, which in some sense
represents the average galactic star formation rate,
But even if the Galaxy were representative of the average,
the disk PDMF does not include the halo star formation.
Thus, one should correct the CSFR contribution to omit
halo star production.  While this is at present impossible to do in detail, 
we can estimate the magnitude of the correction as follows.
The key point is that the low-mass correction is controlled by
the net star formation $\cden$.  Viewed from a Galactic scale, this
is (up to a factor) the total stellar mass today.
But the halo star contribution to the Galactic stellar mass
is small, very probably $\la 10-20\%$ of the total, with the disk
providing the dominant contribution.  Thus, correcting for halo
star formation will raise the low mass points in Figures
\ref{fig:imf} and \ref{fig:alt_imf} by at most a factor of 1.25
(i.e., 0.1 dex in the log).
It is thus unlikely that Population II star formation
contributes significantly to the bimodality seen in the
previous section.  

Another issue one should bear in mind is that the ``local'' star
formation information we have used is that of the PDMF and thus
samples only the solar neighborhood.  It is of course possible that
the shape of the solar neighborhood star formation rate cannot be
simply related to the spatial mean over the disk e.g.,
$\psi_{\odot}(t)/\psi_{\rm disk}(t) \ne const$.  
However, if the shape of $\psi_{\odot}(t)$ is
drastically different from that of $\psi_{\rm disk}(t)$,
this is a problem that plagues not only the present discussion, but
many models of Galactic chemical evolution.  In this case,
the number of
required parameters becomes large, and the strength
of available constraints is thus diluted.

\section{Conclusions} 
\label{sec:fin}

It is both simple and conventional to assume that
the Galaxy's star formation history:
\begin{enumerate}
\item is typical of the universal average (i.e., $\psi \propto \csfr$), \\
\item has an IMF that is constant in time, and  \\
\item has an IMF that is smooth, i.e., unimodal.
\end{enumerate}
In this paper, we have tested the compatibility
of these assumptions.  
We used the observed PDMF to derive the IMF implied by 
star formation universality and IMF constancy 
(i.e., assumptions 1 and 2).  The resulting IMF is consistent
with unimodal in a formal statistical sense, but shows
clear signs of bimodality,
with a low mass peak at $m \sim 0.2 \msol$, and
a second peak around $m \sim 1.3 \msol$ which we suspect is unphysical.
If the apparent bimodality persists and is strengthened
as the data improve, 
then the IMF is not in accord with assumption 3, and
we conclude that one of the three premises is false.

Which assumption fails?
If star formation universality does not hold, 
it could be for many reasons.
For example, the disk star formation history
could differ from the full Galactic history (i.e., 
there may have been much star formation in halo,
as discussed in \S\ref{sec:local}).
Or it could well be that the CSFR is dominated, at 
least at high-redshift, by galaxies with a very
different history than our own.
Alternatively, it is possible that our Galaxy's evolution
was typical, and one {\em can} use the
CSFR for Galactic evolution purposes

If the Galaxy's star formation is indeed representative,
then one of our other two assumptions must fail.
For example, the IMF could be a time-varying,
and biased towards high masses at early epochs; this
is certainly possible but not conventional.
Or it is possible that the universal IMF is indeed
bimodal, as in Figure \ref{fig:imf}.
However, for our purposes it is not encouraging that there is no hint of
bimodality in the PDMF, and it is perhaps suspicious
that the bimodality in the IMF
we derive sets in just at the range of masses where the
IMF-PDMF transformation is nontrivial.

The test we have described can be considerably sharpened as
the observations, both cosmic and local, improve.  
Most importantly, a better understanding of extinction properties at high
redshift will help nail down the CSFR.  
The PDMF errors are large, could also be improved
with additional data over the whole mass range, but
particularly in the region where the bimodality of 
Figure \ref{fig:imf} appears,
i.e., $m \simeq 0.8 - 2 \msol$, corresponding to 
A, F, and G stars.
Indeed, to our knowledge the PDMF has not been revisited
since Scalo \pcite{scalo}; 
a full re-analysis would be of great usefulness.

Other observations and analyses can help to determine which assumptions
hold and which do not.  Regarding the IMF, there
is observational evidence, in elemental ratios, that
the IMF does not vary much in space and time (at least
at the high mass end, Wyse \cite{rosie}).
Direct determination of the IMF in local star forming regions
is difficult; it is nevertheless tantalizing that a recent analysis
of the stellar population of the Orion Trapezium
(Hillenbrand \cite{hill})
has found a mass function
broadly consistent with a Miller-Scalo \pcite{ms}
form, showing no bimodality.
Further tests of cosmic star formation
are also possible, e.g., using the CSFR to reproduce the
classic chemical evolution results for various
systems and epochs (Fall \& Pei \cite{fp};
Cass\'e, Olive, Vangioni-Flam, \& Audouze \cite{cova}).
Finally, it is of great interest to use local
observables to infer the Galactic star formation history.
One such method combines the local data on the  G-dwarf and
age-metallicity distributions 
(e.g. Rocha-Pinto \& Maciel \cite{rpm}, and references therein).  
This technique remains uncertain due to the difficulties in
obtaining an accurate age-metallicity relation, particularly 
for large ages.  Nevertheless, it is noteworthy that
the derived LSFR, while increasing towards the past, may not
show as strong an increase as the cosmic rate at early epochs.
Clearly, new and improved tracers of the LSFR would be useful.

Finally, the most likely conclusion is perhaps that 
the Milky Way is not typical of the average galaxy over all epochs.
If so, it would be interesting to determine if the Milky Way evolution
is at least typical of all spiral galaxies.  
In any case, we already see hints that the simplest of
all possible worlds---in which assumptions 1, 2, and 3 hold---is
not the one we live in.  Chemical evolutionists take note.

\acknowledgments
I am pleased to acknowledge useful
conversations with Dave Graff, Katie Freese, 
Grant Mathews, Keith Olive,
and Evan Skillman.  I thank the anonymous referee, and 
the editor
Steve Shore, for helpful suggestions which improved this paper.
This work was supported in part by
DoE grant DE-FG02-94ER-40823.

\nobreak

\newpage

\centerline{\large \bf Figure Captions}

\begin{enumerate}
\item
\label{fig:imf}
The IMF derived from the PDMF of Scalo \pcite{scalo} and 
the Madau fiducial, ``constant $E_{B-V}$'' CSFR ({\em inset}).
IMF units are arbitrary.  

\item
\label{fig:quad_fit}
(a) The IMF of Figure \ref{fig:imf}, fit to a lognormal via
$\chi^2$ minimization.  \\
(b) The residuals (i.e., $\log \phi - \log \phi_{\rm fit}$) of
panel (a).  The discontinuity in the $1-2 \, \msol$ range 
suggests that the points in this range do not vary randomly
but instead show to a real departure from the simple unimodal fit.

\item
\label{fig:alt_imf}
The IMF plotted as in Fig.\ \ref{fig:imf} for different CSFR forms,
as noted in the inset.  
{\em Filled circles}: fiducial Madau CSFR of Fig.\ \ref{fig:imf};
{\em open triangles}: ``evolving $E_{B-V}$'' Madau CSFR ;
{\em open squares}: ``truncated'' Madau CSFR.
Errorbars have been suppressed for clarity.

\item
\label{fig:pdmf_Salp}
The PDMF, plotted as $m \pdmf(m)$;
{\em filled circles}:  observational data are from Scalo \pcite{scalo}.
The PDMF is derived using
the Madau CSFR models as indicated in the inset
and assuming an 
IMF of 
the Salpeter \pcite{salp} power law form, $\phi(m) \propto m^{-2.35}$.

\item
\label{fig:pdmf_MS}
The PDMF as in Figure \ref{fig:pdmf_Salp}, assuming an IMF of 
the Miller-Scalo \pcite{ms} lognormal
form.
\end{enumerate}

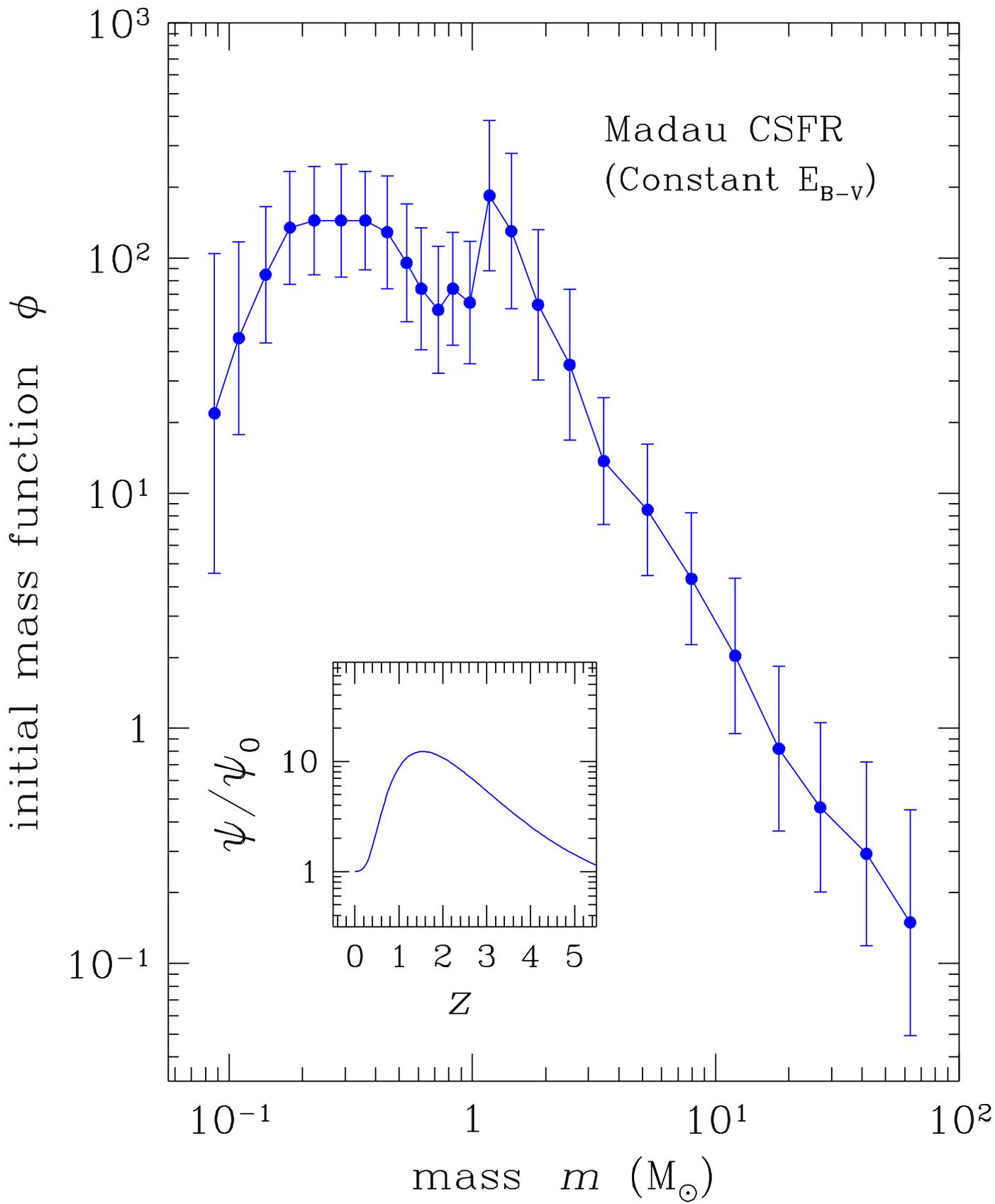
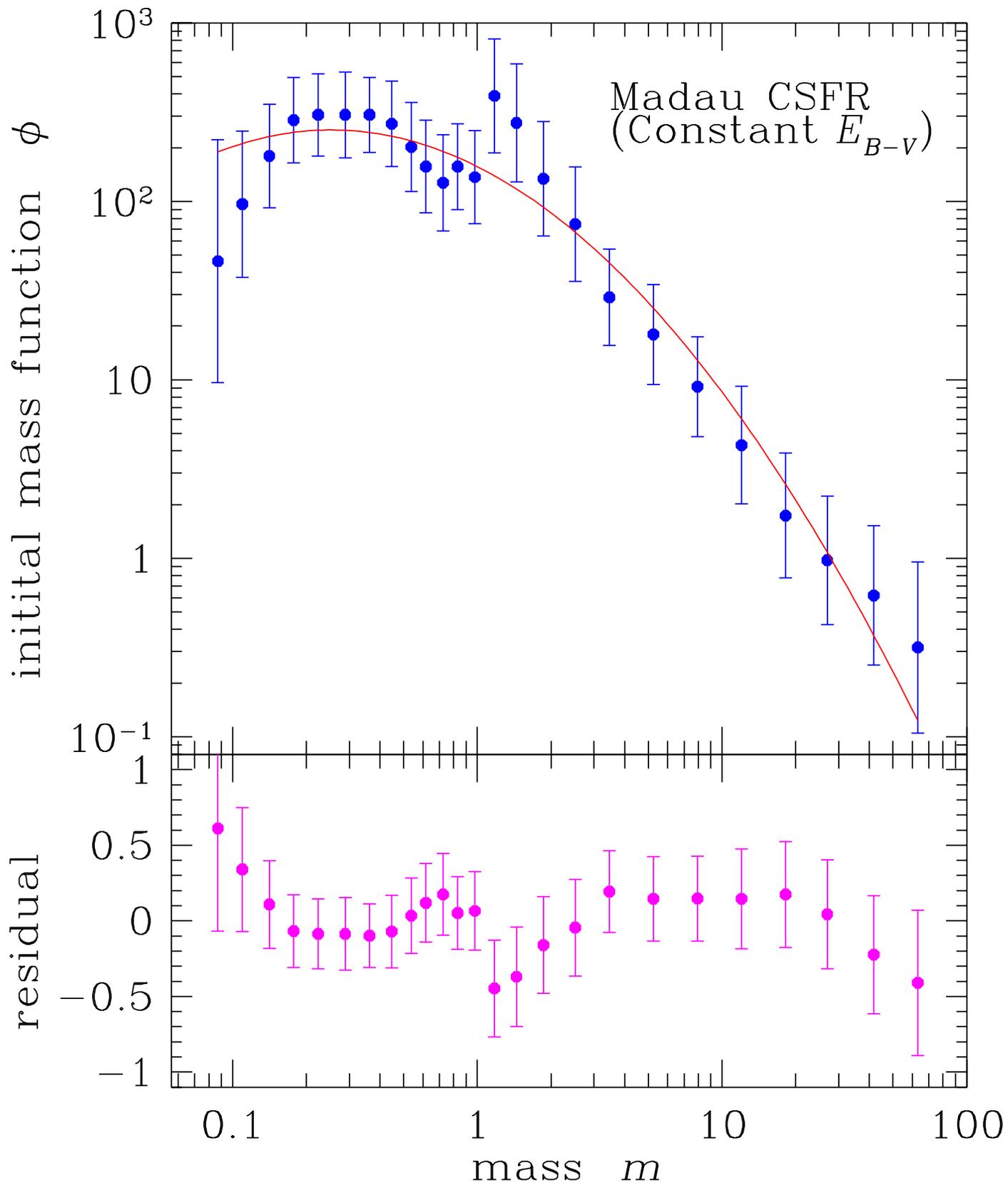
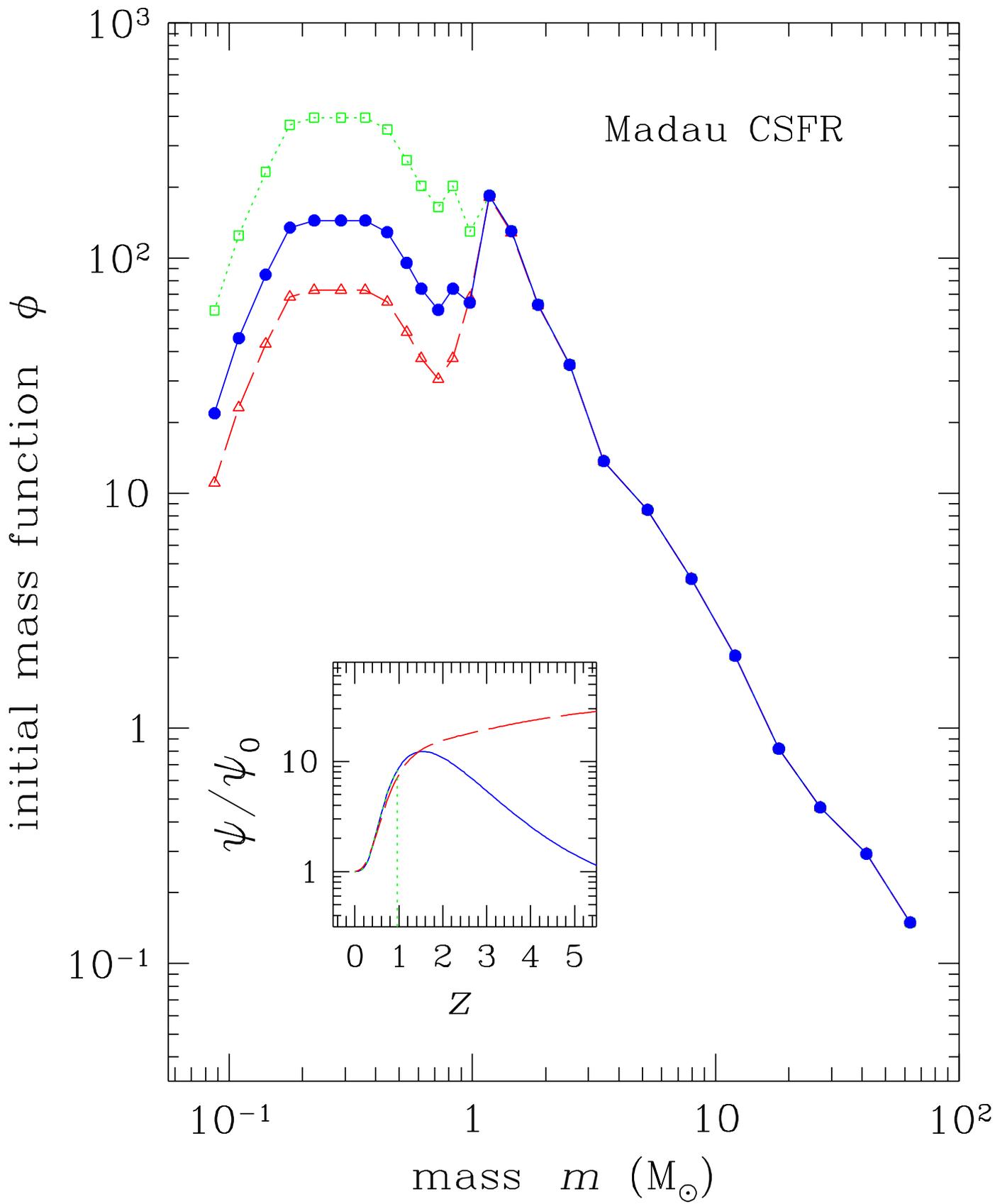
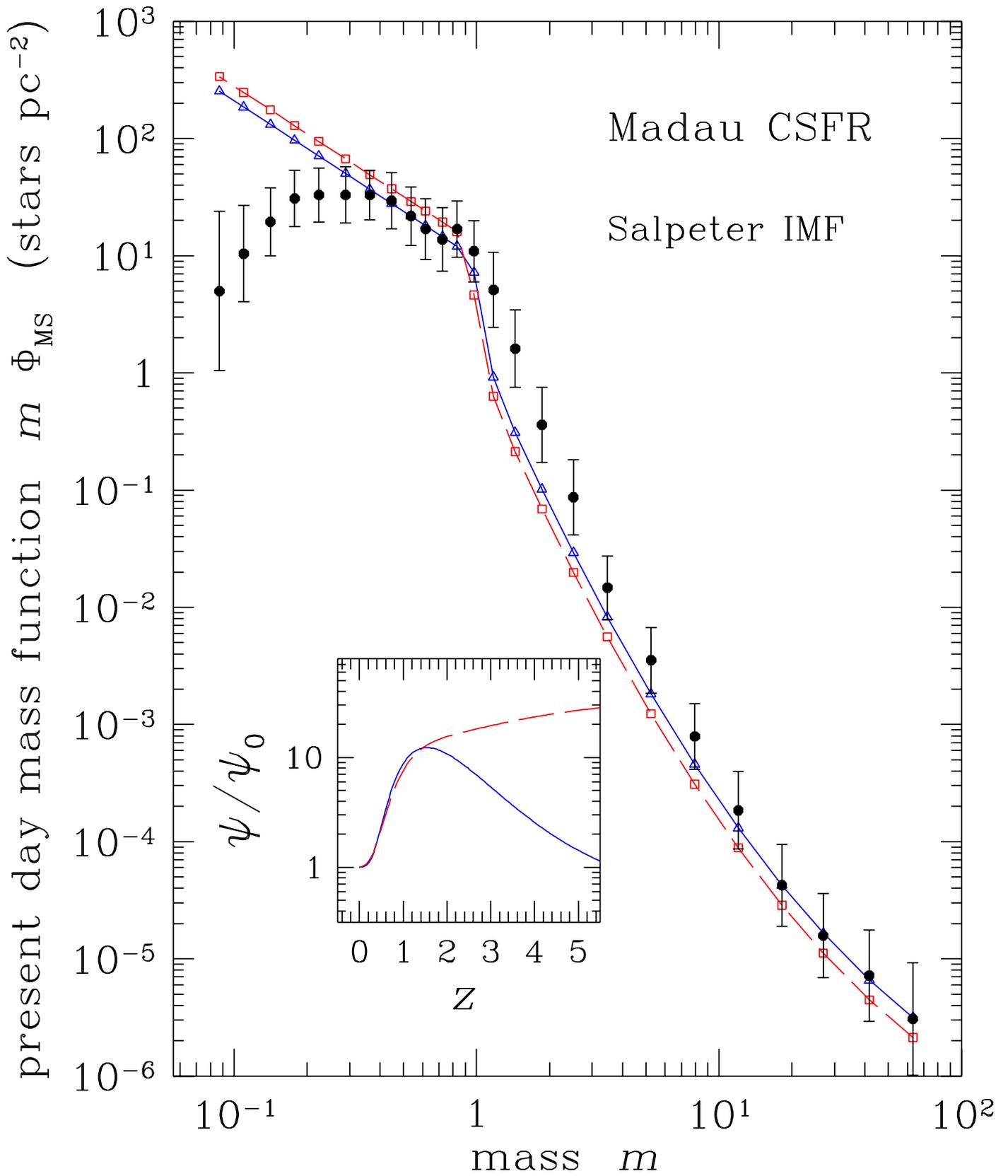
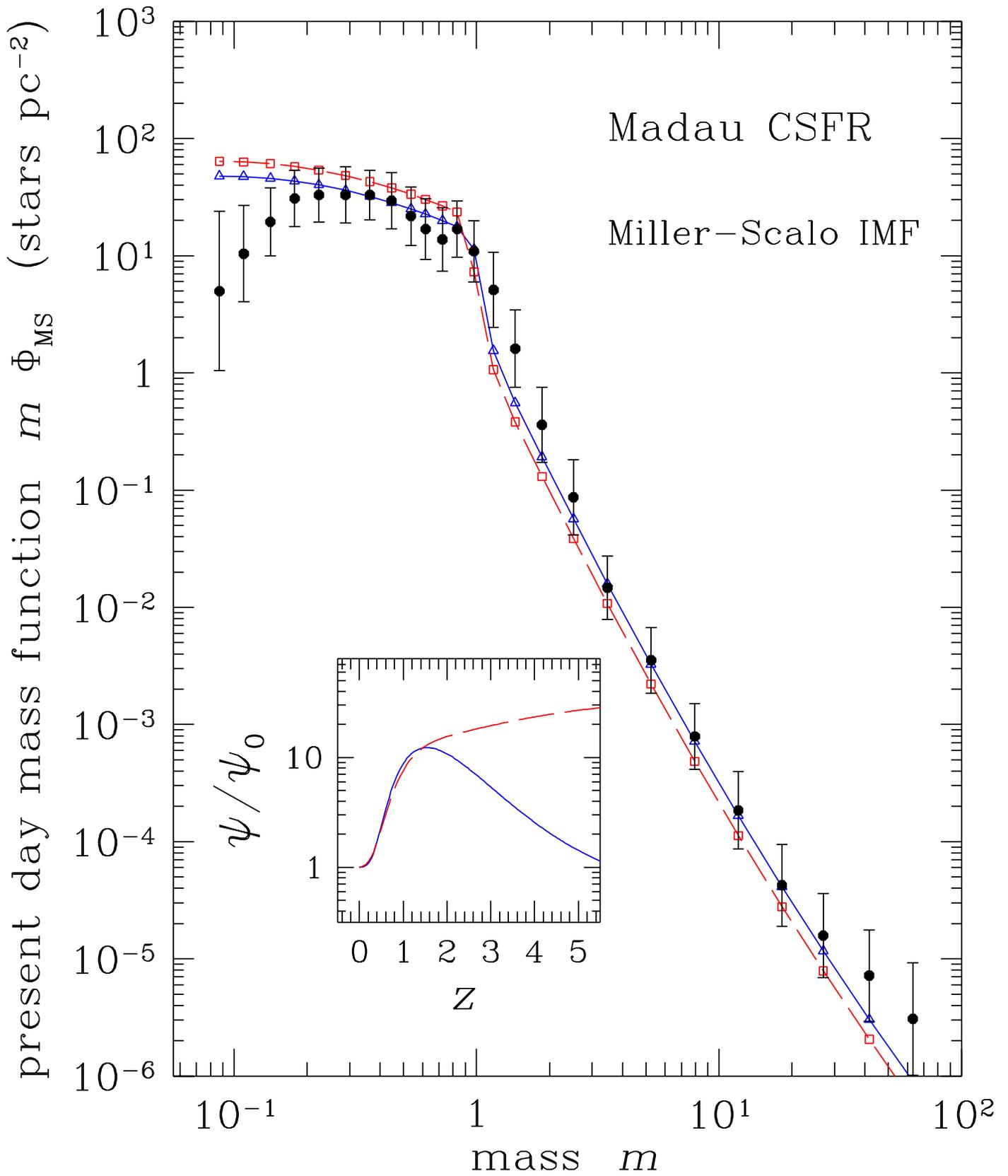
\end{document}